\documentclass[12pt,preprint]{aastex}

\begin{document}
\title{X-Ray and high energy flares from late internal shocks of gamma-ray bursts}
\author{Y. W. Yu$^{1,2}$ and Z. G. Dai$^1$}
\affil{$^1$Department of Astronomy, Nanjing University, Nanjing 210093, China; \\yuyw, dzg@nju.edu.cn
\\$^2$Institute of Astrophysics, Huazhong Normal University, Wuhan
430079, China }
\begin{abstract}
We study afterglow flares of gamma-ray bursts (GRBs) in the
framework of the late internal shock (LIS) model based on a careful
description for the dynamics of a pair of shocks generated by a
collision between two homogeneous shells,. First, by confronting the
model with some fundamental observational features of X-ray flares,
we find some constraints on the properties of the pre-collision
shells that are directly determined by the central engine of GRBs.
Second, high energy emission associated with X-ray flares, which
arises from synchrotron self-Compton (SSC) emission of LISs, is
investigated in a wide parameter space. The predicted flux of high
energy flares may reach as high as $\sim 10^{-8}\rm
erg~cm^{-2}s^{-1}$, which is likely to be detectable with the Large
Area Telescope (LAT) aboard \textit{the Fermi Space Telescope}
(formerly GLAST).
\end{abstract}

\section{Introduction}
Since the launch of the \textit{Swift} satellite in 2004, the X-ray
telescope (XRT) aboard has revealed several new features of early
X-ray afterglow light curves of gamma-ray bursts (GRBs) (e.g.,
M\'ez\'aros 2006; Zhang 2007 for reviews): (i) the transition from
prompt phase to afterglow phase usually exhibits steep decline of
X-ray flux, which is generally interpreted as the high-latitude tail
of prompt emission (i.e., curvature effect; Fenimore et al. 1996;
Kumar \& Panaitescu 2000;), (ii) a shallow decay usually follows the
steep segment during the first hours, which could be caused by a
spread of Lorentz factors in GRB ejecta (Rees \& M\'sez\'aros 1998)
or a continuous energy injection into GRB ejeta due to lasting
activities of GRB central engines (e.g., Dai \& Lu 1998a, 1998b;
Zhang \& M\'esz\'aros 2001a; Dai 2004; Fan \& Xu 2006; Yu \& Dai
2007), and (iii) bright X-ray flares superimposing on underlying
afterglow emission were observed from nearly a half of the
\textit{Swift} GRBs (Burrows et al. 2005; see Chincarini et al. 2007
and Falcone et al. 2007 for more recent statistic studies). These
discoveries are of great importance for revealing the nature of GRB
central engines, and especially, X-ray flares are widely accepted to
be due to some delayed intermittent activities of the central
engines.

In the sight of model, the delayed intermittent activities might be
episodic accretion onto a central object due to a chopped accretion
disk (Perna et al. 2006), or episodic accretion due to a modulation
of the accretion flow by a magnetic barrier (Proga \& Zhang 2006),
or magnetic reconnection on a nascent differentially-rotating
massive neutron star (Dai et al. 2006), etc. Many scenarios were
designed in the past few years, but none of them is conclusive (see
Zhang 2007 for a review). However, the key point is that, in almost
all of these scenarios, internal dissipations in late-ejected
materials are usually required to produce flare emission. The
leading mechanism for internal dissipation is late internal shocks
(LISs; Burrows et al. 2005; Fan \& Wei 2005), which are generated by
collisions among some shells ejected at different times with
different Lorentz factors and energies. Hitherto, the properties of
LIS-produced emission have been discussed for a few times (e.g., Fan
\& Wei 2005; Wu et al. 2005; Zhang et al. 2006; Zou et al. 2006; Fan
et al. 2007; Galli \& Guetta 2008). However, the properties of the
pre-collision shells, which are directly determined by the late
activities of the central engines, were not concerned in their
works. In this paper, in contrast, we give a more detailed
description for the dynamics of LISs including internal forward and
reverse shocks, and then investigate the predicted X-ray flare
luminosity and the shape of light curves within a wide model
parameter space. By confronting the theoretical predictions with
some observational features of X-ray flares, we can find some
constraints on the model parameters.

Furthermore, it must be helpful to constrain the model more
stringently by performing a simultaneous observation in the
high-energy $\gamma$-ray bands for X-ray flares in the
\textit{Fermi} era. As suggested previously by Wang, Li, \&
M\'esz\'aros (2006), Fan \& Piran (2006), and Galli \& Guetta
(2008), high energy flares are expected to arise from inverse
Compton (IC) scattering of low-energy flare photons off some
relativistic electrons. These electrons belong to either GRB ejecta
(i.e. external inverse Compton, EIC) or flare ejecta self (i.e.
synchrotron self-Compton, SSC). Based on the same consideration, we
discuss the detectability of high energy flares in the
\textit{Fermi} era with the careful description for LIS dynamics
along with the constraints from X-ray observations.

This paper is organized as follows: in section 2, we describe the
dynamics of a pair of shocks arising from a collision between two
homogeneous shells and the resulting synchrotron radiation of the
shocked electrons. In section 3, the model is tested by some
observational features of X-ray flares. In section 4, we estimate
the accompanying SSC high-energy emission and its detectability in
the \textit{Fermi} era. Finally, a summary is given in section 5.

\section{The model}
After the prompt phase of a GRB, the central engine might still be
able to eject some separate shells with different Lorentz factors
and energies due to its delayed intermittent activities. Collisions
between these shells generate LISs, which give rise to the observed
X-ray flares. For simplicity, the scope of the paper is restricted
within one collision between two homogeneous shells.

\subsection{Dynamics}
We set the time zero point at the GRB trigger and measure time in
the observer's frame. At a time of $t_{\rm ej,A}$, the central
engine ejects a shell denoted by $A$, which moves at a constant bulk
Lorentz factor $\gamma_{\rm A}$ and carries an isotropic-equivalent
kinetic energy luminosity $L_{\rm iso,A}$. The particle number
density of the shell measured in its comoving frame (denoted by a
prime hereafter) can be calculated by
\begin{equation}
n'_{\rm A}={L_{\rm iso,A}\over 4\pi R^2\gamma_{\rm A}^2m_pc^3},\label{den}
\end{equation}
where $R$ is the radius of the shell. Some time ($\Delta t_{\rm
ej}$) later, another shell denoted by $B$ with $\gamma_{\rm B}$ and
$L_{\rm iso,B}$ is assumed to be ejected again. We require
$\gamma_{\rm B}>\gamma_{\rm A}$ in order to let shell $B$ catch up
and collide with the prior shell $A$. Consequently, a collision
between $A$ and $B$ takes place at the radius $R_{\rm
col}={\beta_{\rm A}\beta_{\rm B}c\Delta t_{\rm ej}/(\beta_{\rm
B}-\beta_{\rm A}})$ and the time $t_{\rm col}=t_{\rm
ej,A}+{\beta_{\rm A}\Delta t_{\rm ej}/(\beta_{\rm B}-\beta_{\rm
A}})-{R_{\rm col}/c}$. For $(\gamma_{\rm A},\gamma_{\rm B})\gg1$,
they read
\begin{equation}
R_{\rm col}\simeq{2\gamma_{\rm A}^2c\Delta t_{\rm ej}\over1-(\gamma_{\rm A}/\gamma_{\rm B})^2},
\end{equation}
\begin{equation}
t_{\rm col}\simeq t_{\rm ej,A}+{\Delta t_{\rm ej}\over1-(\gamma_{\rm
A}/\gamma_{\rm B})^2}\simeq t_{\rm onset}.
\end{equation}
The collision time $t_{\rm col}$ determines the observed onset time
$t_{\rm onset}$ of a flare in physics. Strictly, $t_{\rm onset}$
should be mildly larger than $t_{\rm col}$ since the physical onset
is usually buried by normal afterglow emission.

Due to the collision, a pair of shocks are generated, including a
forward shock and a reverse shock that propagate into shells $A$ and
$B$, respectively. Separated by the two shocks and a contact
discontinuity surface, the system is divided into four regions: (1)
unshocked shell $A$, (2) shocked shell $A$, (3) shocked shell $B$,
and (4) unshocked shell $B$, bulk Lorentz factors of which are
$\gamma_1=\gamma_{\rm A}$, $\gamma_2=\gamma_3\equiv\gamma$, and
$\gamma_4=\gamma_{\rm B}$. Considering the motion of the shocked
regions relative to unshocked regions 1 and 4, respectively, two
relative Lorentz factors can be calculated by
\begin{equation}
\gamma_{21}={1\over2}\left({\gamma_1\over\gamma}+{\gamma\over\gamma_1}\right),~~
\gamma_{34}={1\over2}\left({\gamma\over\gamma_4}+{\gamma_4\over\gamma}\right).\label{relgam}
\end{equation}
Then, according to the jump conditions between the two sides of a
shock (Blandford \& McKee 1976), we can calculate the internal
energy densities of the two shocked regions by
$e'_2=(\gamma_{21}-1)(4\gamma_{21}+3)n'_1m_pc^2$ and
$e'_3=(\gamma_{34}-1)(4\gamma_{34}+3)n'_4m_pc^2$, where $n'_1=n'_A$
and $n'_4=n'_B$. The mechanical equilibrium between the two shocked
regions requires $e'_2=e'_3$, which yields
\begin{equation}
{(\gamma_{21}-1)(4\gamma_{21}+3)\over(\gamma_{34}-1)(4\gamma_{34}+3)}={n'_4\over n'_1}=\left({L_4\over
L_1}\right)\left({\gamma_1\over \gamma_4}\right)^2\equiv f.\label{dyne}
\end{equation}
For four given parameters, $L_1,~L_4,~\gamma_1,$ and $\gamma_4$,
that describe the basic properties of the pre-collision shells, we
can obtain the value of $\gamma$ by solving equations (\ref{relgam})
and (\ref{dyne}). In particular, for four limits shown in the
($\gamma_4/\gamma_1$, $L_4/L_1$) parameter space (see Figure 1),
these equations can be solved analytically. For
$\gamma_4\gg\gamma_1$, we have (i) if ${L_4/
L_1}\gg{(1/7)}\left({\gamma_4/\gamma_1}\right)^4$,
\begin{equation}
\gamma_{21}={\gamma_4\over2\gamma_1}\gg1~ \&~
\gamma_{34}-1\approx{\gamma_4^2\over7f\gamma_1^2}=\xi\ll1;~~\gamma=\gamma_4(1-\sqrt{2\xi}),
\end{equation}
which means the forward shock is relativistic and the reverse shock
is Newtonian (RFS \& NRS); (ii) if $16\ll{L_4/
L_1}\ll{(1/16)}\left({\gamma_4/\gamma_1}\right)^4$,
\begin{equation}
\gamma_{21}={f^{1/4}\gamma_4^{1/2}\over2\gamma_1^{1/2}}\gg1~ \&~
\gamma_{34}={\gamma_4^{1/2}\over2f^{1/4}\gamma_1^{1/2}}\gg1;~~\gamma=f^{1/4}\gamma_1^{1/2}\gamma_4^{1/2},
\end{equation}
which means both the forward and reverse shocks are relativistic
(RFS \& RRS); (iii) if ${L_4/ L_1}\ll7$,
\begin{equation}
\gamma_{21}-1\approx{f\gamma_4^2\over7\gamma_1^2}=\xi\ll1~ \&~
\gamma_{34}={\gamma_4\over2\gamma_1}\gg1;~~\gamma=\gamma_1(1+\sqrt{2\xi}),
\end{equation}
which means the forward shock is Newtonian whereas the reverse shock
is relativistic (NFS \& RRS). Finally, (iv) for
$\gamma_4\approx\gamma_1$, both the forward and reverse shocks are
Newtonian (NFS \& NRS). Since $\gamma_1$, $\gamma_4$, and the ratio
$f$ are unchanged with the moving of the shells, the values of
$\gamma$ as well as $\gamma_{21}$ and $\gamma_{34}$ can keep
constant before the shocks cross the shells. In principle, when one
shock crosses the corresponding shell, the other shock should be
decelerated in the rest frame of its upstream material. However, if
the energy carried by the remaining unshocked material is much less
than the energy of the total shocked material at that time, $\gamma$
would be not changed significantly. So, we can treat with $\gamma$
as a constant always approximately if we assume the crossing times
of the two shocks to be equal more or less. After both shocks
vanish, the merged shell no longer interacts with any other
materials until it is caught up with by the third late-ejected
shell.

The above analysis on kinematics and dynamics yields the evolution
of the radius of the system after the collision,
\begin{equation}
R(t)=R_{\rm col}+2\gamma^2c(t-t_{\rm col})\equiv R_{\rm col}+2\gamma^2cT,\label{Rt}
\end{equation}
where we redefine the time $T$ by resetting the time zero point from
the GRB trigger to the flare onset time. We can define an initial
expansion time from equation (\ref{Rt}),
\begin{equation}
T_{\rm exp}={R_{\rm col}\over2\gamma^2c}={\Delta t_{\rm
ej}\over1-(\gamma_1/\gamma_4)^2}\left({\gamma_1\over\gamma}\right)^2\label{texp},
\end{equation}
and find that the increase of the radius before $T_{\rm exp}$ can be
ignored (i.e. $R\simeq\rm constant$) but after $T_{\rm exp}$ the
radius increases linearly with time (i.e. $R\propto T$). Moreover,
as the propagation of the shocks, the total electron numbers can be
calculated by $N_{e,2}=8\pi R^2 n'_{1}({\gamma_{\rm 21}\beta_{\rm
21}/\gamma\beta})\gamma^2cT$ and $N_{e,3}=8\pi R^2
n'_{4}({\gamma_{\rm 34}\beta_{\rm 34}/\gamma\beta})\gamma^2cT$ for
the two shocked regions (Dai \& Lu 2002),
both of which are proportional to $T$ before the shock crossing.

\subsection{Synchrotron radiation}
For the two shocked regions, fractions $\epsilon_{B}$ and
$\epsilon_{e}$ of the internal energy are assumed to be carried by
magnetic fields and hot electrons, respectively. Then, the strength
of the magnetic fields reads $B'_i=(8\pi \epsilon_{B}e'_i)^{1/2}$,
whose variation is determined by the evolution of $e'_i$. Denoting
the comoving volume of a shocked region as $V'_i$, we adopt
$e'_i\propto {V'}_i^{-1}\propto R^{-2}$ by ignoring a possible
spreading of the hot shocked materials. In the presence of the
shocks that might be able to suppress the spreading of the hot
materials, the neglect of the spreading may be plausible. However,
after the shock crossing time $T_{\rm crs}$ (we assume here the two
shocks cross at a similar time), the spreading of the hot materials
into the vacuum cannot be ignored and the hot materials should
experience an adiabatic cooling. During this phase, we assume that
the volume of the merged shell is determined by a simple power-law
as $V'_i\propto R^s$, where $s$ is a free parameter and its value is
taken from 2 to 3. Then, the particle number densities would
decrease as $n'_i\propto {V'}_i^{-1}\propto R^{-s}$ and the internal
energy densities as $e'_i\propto {V'}_i^{-4/3}\propto R^{-4s/3}$ due
to adiabatic cooling. Therefore, a multi-power temporal behavior for
the magnetic field strength can be found:
\begin{equation}
B'_{i}\propto\left\{\begin{array}{ll} T^0,~~~~~~~T<T_{\rm
exp};\\
T^{-2s/3},~~T>T_{\rm exp};
\end{array}\right.~~~~~T_{\rm crs}<T_{\rm exp},
\end{equation}
\begin{equation}
B'_{i}\propto\left\{\begin{array}{ll} T^0,~~~~~~~T<T_{\rm
exp};\\T^{-1},~~~~~T_{\rm exp}<T<T_{\rm crs};\\
T^{-2s/3},~~T>T_{\rm crs};
\end{array}\right.~~~~~T_{\rm crs}>T_{\rm exp}.
\end{equation}

As usual, we assume that electrons are accelerated by the shocks to
distribute as
$dn'_{e,i}/d{\gamma'}_{e,i}\propto{\gamma'}_{e,i}^{-p}$ for
${\gamma'}_{e,i}>{\gamma'}_{e,m,i}$, where the minimum Lorentz
factor is defined as $\gamma'_{e,m,i}=\epsilon_{e}C_p({m_p/
m_e})(\gamma_{\rm rel}-1)$ with $C_p\equiv(p-2)/(p-1)$ and
$\gamma_{\rm rel}=\gamma_{21}$ or $\gamma_{34}$.
Additionally, the cooling Lorentz factor, ${\gamma'}_{e,c,i}=6\pi
m_ec/[ (1+Y_i)\sigma_T{B'}_i^2\gamma T]$, also needs to be
introduced, above which the electrons lose most of their enegies.
The Compton parameter $Y_i$ defined as the ratio of the IC to
synchrotron luminosities can be estimated by $Y_i\approx
[(4\eta_i\epsilon_{e}/\epsilon_{B}+1)^{1/2}-1]/2$ with
$\eta_i=\min[1,(\gamma'_{e,c,i}/\gamma'_{e,m,i})^{2-p}]$ (Sari \&
Esin 2001). Then, by calculating two characteristic frequencies and
a peak flux density,
\begin{equation}
\nu_{m,i}={ q_e\over2\pi m_ec}{B'}_i{\gamma'}_{e,m,i}^2\gamma,~~\nu_{c,i}={ q_e\over2\pi
m_ec}{B'}_i{\gamma'}_{e,c,i}^2\gamma,~~F_{\nu,\max,i}={ N_{e,i}\over4\pi d_{L}^2}{ m_{e}c^2\sigma_{T}\over
3q_e}{B'}_{i}\gamma,
\end{equation}
where $d_L$ is the luminosity distance of the source, a multi-power
synchrotron spectrum contributed by a shock can be written as (Sari
et al. 1998)
\begin{equation}
F_{\nu,i}=F_{\nu,\max,i}\times\left\{
\begin{array}{ll}
\left({\nu\over\nu_l}\right)^{1/3},~~~~~~~~~~~~~~~~~~~~~~\nu<\nu_l;\\
\left({\nu\over\nu_l}\right)^{-(q-1)/2},~~~~~~~~~~~~~~~~\nu_l<\nu<\nu_h;\\
\left({\nu_h\over\nu_l}\right)^{-(q-1)/2}\left({\nu\over\nu_h}\right)^{-p/2},~~~~\nu_h<\nu,\\
\end{array}\right.
\end{equation}
where $\nu_{l}=\min(\nu_{m,i},\nu_{c,i})$,
$\nu_{h}=\max(\nu_{m,i},\nu_{c,i})$, and $q=2$ for
$\nu_{c,i}<\nu_{m,i}$ and $q=p$  for $\nu_{c,i}>\nu_{m,i}$.

To obtain X-ray light curves further, the temporal dependence of the
characteristic quantities also needs to be presented as follows:

(i) In the case of $T_{\rm crs}<T_{\rm exp}$,
\begin{equation}
\nu_{m,i}\propto\left\{\begin{array}{ll} T^0,~~~~~T<T_{\rm exp};\\T^{-2s/3},~~T>T_{\rm
exp},\end{array}\right.\label{num}
\end{equation}
\begin{equation}
\nu_{c,i}\propto\left\{\begin{array}{ll} T^{-2},~~~~T<T_{\rm exp};\\T^{2s-2},~~T>T_{\rm
exp},\end{array}\right.\label{nuc}
\end{equation}
\begin{eqnarray}
F_{\nu,\max,i}\propto\left\{\begin{array}{ll} T,~~~~~T<T_{\rm
crs};\\
T^0,~~~~~T_{\rm crs}<T<T_{\rm exp};\\
T^{-2s/3},~~T>T_{\rm exp};\end{array}\right. \label{fmax}
\end{eqnarray}

(ii) In the case of $T_{\rm crs}>T_{\rm exp}$,
\begin{equation}
\nu_{m,i}\propto\left\{\begin{array}{ll} T^0,~~~~~T<T_{\rm
exp};\\T^{-1},~~~~T_{\rm exp}<T<T_{\rm crs};\\
T^{-2s/3},~~T>T_{\rm crs},
\end{array}\right.\label{num}
\end{equation}
\begin{equation}
\nu_{c,i}\propto\left\{\begin{array}{ll} T^{-2},~~~~T<T_{\rm
exp};\\T,~~~~~T_{\rm exp}<T<T_{\rm crs};\\
T^{2s-2},~~T>T_{\rm crs},
\end{array}\right. \label{nuc}
\end{equation}
\begin{eqnarray}
F_{\nu,\max,i}\propto\left\{\begin{array}{ll} T,~~~~~T<T_{\rm exp};\\
T^0,~~~~~T_{\rm exp}<T<T_{\rm crs};\\
T^{-2s/3},~~T>T_{\rm crs}.\end{array}\right. \label{fmax}
\end{eqnarray}
From the above expressions, we can see that $\nu_{c,i}$ reaches its
minimum value at $T_{\rm exp}$, while $\nu_{m,i}$ starts to decrease
at the same time. So, the relationships between $\nu_{c,i}$ and
$\nu_{m,i}$ as well as the spectra and light curves can be found
easily by fixing and comparing the values of $\nu_{c,i}$ and
$\nu_{m,i}$ at $T_{\rm exp}$.

\section{X-ray flares}
Observations have shown that X-ray flares may consist of one or a
few pulses. We consider one pulse to be mainly produced by one
collision between two shells. Here we do not try to fit X-ray
observational data in detail, which requires a more elaborate model
that takes the fine structure of pre-collision shells into account.
Instead, we only test the model by some fundamental observational
features of X-ray flares.

\subsection{X-ray luminosity}
Statistic studies found that the average flare fluence (in $0.2-10$
keV band) is approximately a factor of ten less than the fluence of
prompt emission as $\sim 10^{-7}~\rm erg~cm^{-2}$ (Falcone et al.
2007) and the flare peak times $t_{\rm peak}$ concentrate into the
range from 100 to 1000 s (Chincarini et al. 2007). Moreover,
according to $\delta t/t_{\rm peak}\ll1$, the temporal width of
flares, $\delta t$, can be estimated to be from a few to several ten
seconds. By considering the spread of the distributions of these
quantities, we suggest to take $\sim10^{-9}~\rm erg~s^{-1}~cm^{-2}$
as the lower limit for the peak flux of X-ray flares. For a typical
luminosity distance $d_{L}=10^{28}$ cm for GRBs, we give
$\sim10^{48}~\rm erg~s^{-1}$ as the lower limit for X-ray flare
luminosity.

In order to derive some constraints on the model from this
luminosity lower limit, first we show two ($\gamma_1,\gamma_4$)
parameter spaces in Figure 2 with certain values of $L_1$ and $L_4$:
$L_4/L_1\gg1$ for the left panel and $L_4/L_1=1$ for the right
panel, which correspond to different dynamic cases. According to the
variation of $\nu_{m}^*$ and $\nu_c^*$,\footnote{The star
superscription represents that the values are calculated at the time
of $T_{\rm exp}$. We obtain the values of $\nu_{m}^*$ and $\nu_c^*$
from the spectrum due to the shock that dominates the X-ray emission
and thus the subscription $i$ of the quantities is omitted in this
section.} the parameter spaces can be roughly divided into four
regions denoted by `a, b, c, d', where different relationships among
$\nu_{m}^*$, $\nu_c^*$, and $\nu_X(\equiv2.4\times10^{17}\rm Hz)$
are given as listed in Table 1. We then use some shaded contours to
show the regions where model-predicted X-ray luminosity exceeds the
observational lower limit. Obviously, region `a' is restricted
significantly by this luminosity constraint due to the slow cooling
of electrons (as indicated by $\nu_{m}^*<\nu_{X}<\nu_{c}^*$) that
leads to a low radiation efficiency. To be more specific, we can
conclude further that (i) in order to produce sufficiently strong
flare emission, it is required that $\gamma_4>{\rm
few}\times\gamma_1$. This means that at least one of the forward and
reverse shocks is mild-relativistic; and (ii) when
$\gamma_4\approx{\rm few}\times\gamma_1^2$, we can obtain the
highest theoretical X-ray luminosity, because this parameter region
roughly locates the boundary between regions `c' and `d', where the
relationship $\nu_{c}^*<\nu_X\sim\nu_{m}^*$ can be found, indicating
a high X-ray radiation efficiency. Second, in order to uncover the
dependence of the emission on parameters $L_1$ and $L_4$, we also
show two corresponding parameter spaces in Figure 3 with certain
values of $\gamma_1$ and $\gamma_4$. A minimum value of
$\sim10^{50}-10^{51}~\rm erg~s^{-1}$ for $L_4$ can be found, but the
constraint on $L_1$ is loose. This indicates that the resulting
X-ray luminosity is mainly determined by the kinetic-energy
luminosity of the lagged rapid shell rather than that of the leading
slow shell.

Finally, in the above calculations for X-ray luminosity, we ignored
possible synchrotron self-absorption of the X-ray photons, which is
able to suppress the X-ray emission. The synchrotron self-absorption
thickness at the X-ray band can be calculated by (Panaitescu \&
Kumar 2000)
\begin{equation}
\tau_{ssa,\nu_{X}}\simeq{5q_eN_e\over4\pi R^2B'\gamma_{e,l}^{'5}}\times\left\{
\begin{array}{ll}
\left({\nu_X\over\nu_l}\right)^{-5/3},~~~~~~~~~~~~~~~~~~~~~~~~~\nu_X<\nu_l;\\
\left({\nu_X\over\nu_l}\right)^{-(q+4)/2},~~~~~~~~~~~~~~~~~~~~~\nu_l<\nu_X<\nu_h;\\
\left({\nu_h\over\nu_l}\right)^{-(q+4)/2}\left({\nu_X\over\nu_h}\right)^{-(p+5)/2},~~~~\nu_h<\nu_X,\\
\end{array}\right.
\end{equation}
where $\nu_{l}=\min(\nu_m,\nu_c)$, $\nu_{h}=\max(\nu_m,\nu_c)$,
$\gamma'_{e,l}=\min(\gamma'_{e,m},\gamma'_{e,c})$, and $q=2$ for
$\nu_{c}<\nu_{m}$ and $q=p$  for $\nu_{c}>\nu_{m}$. By scanning the
whole parameter spaces shown in Figures 2 and 3, we find that the
values of $\tau_{ssa,\nu_{X}}$ are always much lower than unity due
to the large internal shock radius, which means the synchrotron
self-absorption of the X-ray photons can be ignored safely.

\subsection{Shape of X-ray light curves}
We here test the model by considering the shape of the observed
X-ray flare light curves, a basic feature of which is the rapid rise
and fall. Following the frequency relationships given in Table 1, we
can find three possible types of the theoretical X-ray light curves,
as shown schematically by the black lines in Figure 4 for $T_{\rm
crs}\simeq T_{\rm exp}$ where the time zero is set at the flare
onset time. The light curves break at several characteristic times
and the corresponding temporal indices ($\alpha=-d\log F_\nu/d\log
T$) are listed in Table 1. All of the segments after $T_{\rm crs}$
are steepened significantly by the shock crossing effect. By
confronting these theoretical light curves with the observed ones,
on one hand, it is easy to understand the observed steep rise by
resetting the time zero point of the log-log figure from the flare
onset time to the GRB trigger. On the other hand, as discovered by
Liang et al. (2006), the rapid decline of most X-ray flares seems to
be consistent with the curvature effect that predicts a temporal
index being equal to the simultaneous spectral index plus 2. As
shown by the black lines in Figure 4, the intrinsic decline slope of
the last segment is $\alpha=(sp+s)/3$ for all types of the
theoretical light curves and the corresponding spectral index is
$(p-1)/2$ due to $\nu_{m}<\nu_X<\nu_c$. For $s=3$ and $2<p<3$, the
inequation
\begin{eqnarray}
{(sp+s)\over3}>{(p-1)\over2}+2
\end{eqnarray}
can be satisfied easily. This inequation indicates that, in any case
in our model, the observed X-ray flux decay should be dominated by
the curvature effect, as shown by the grey lines in Figure 4. In
addition, although some relatively flat segments with
$\alpha=(3-s)/3$ or $(sp-2s+3)/3$ appear in types II and III light
curves, we still cannot rule out any one of them absolutely, because
these segments that could not be very far from the flare onset time
may be steepened by the time zero effect dramatically.

\section{High energy flares}
Because a part of the synchrotron photons would be boosted to higher
energy by their IC scattering off some relativistic electrons, high
energy counterparts of X-ray flares are expected naturally. In the
following calculations, we restrict our attention within the LIS
model and SSC scenario. Before a specific calculation for the SSC
emission, we first estimate the flux sensitivity of the LAT aboard
the \textit{Fermi} by
\begin{eqnarray}
F_{\rm thr}={5E\over A_{\rm eff}t}=1.33\times10^{-9}\left(E\over{\rm GeV}\right)\left(t\over10^3{\rm
s}\right)^{-1}~\rm erg~cm^{-2}~s^{-1},
\end{eqnarray}
following Zhang \& M\'esz\'aros (2001b) who adopted the criterion
that at least five photons are collected if the instrument is source
dominated, where $E$ is the photon energy and $t$ is the integration
time. The effective area $A_{\rm eff}$ of the instrument is taken as
a constant of $6000~\rm cm^{2}$ and the dependence of the area on
the photon energy is neglected. However, after a transition time of
$\sim2.4\times10^4$s (Gou \& M\'esz\'aros 2007; Yu et al. 2007), the
sensitivity should start to scale as $t^{-1/2}$ due to a limitation
by the background for a long-time observation.

Following Sari \& Esin (2001), the SSC spectrum contributed by a
shock can be obtained by shifting the seed synchrotron spectrum to
higher energy range, i.e. estimating the two break frequencies and
the peak flux density of the SSC spectrum respectively by
\begin{eqnarray}
\nu_{m,i}^{\dag}=2{\gamma'}_{e,m,i}^{2}\nu_{m,i},~\nu_{c,i}^{\dag}=2{\gamma'}_{e,c,i}^{2}\nu_{c,i},~
F_{\nu,\max,i}^{\dag}={\sigma_{T}N_{e,i}\over4\pi R^2}F_{\nu,\max,i}.
\end{eqnarray}
This approximative treatment is finely valid when the effect of the
Klein-Nishina suppression is unimportant. Considering the
highest-energy electrons whose energy enter the Klein-Nishina
regime, we refer to the third break in the SSC spectrum,
\begin{eqnarray}
\nu_{{\rm KN},i}={\gamma^2m_e^2c^4\over h^2\max(\nu_{m,i},\nu_{c,i})},
\end{eqnarray}
above which the SSC spectrum follows $F_{\nu,i}\propto\nu^{-(p-1)}$
(Gupta \& Zhang 2007; Fragile et al. 2004). Therefore, the
approximative SSC spectrum can be summarized as follows (Gupta \&
Zhang 2007)
\begin{equation}
F_{\nu,i}^{\dag}=F_{\nu,\max,i}^{\dag}\times\left\{
\begin{array}{ll}
\left({\nu\over\nu_l}\right)^{1/3},~~~~~~~~~~~~~~~~~~~~~~~~~~~~~~~~~~~~~\nu<\nu_l;\\
\left({\nu\over\nu_l}\right)^{(q-1)/2},~~~~~~~~~~~~~~~~~~~~~~~~~~~~~~~\nu_l<\nu<\nu_h;\\
\left({\nu_h\over\nu_l}\right)^{-(q-1)/2}\left({\nu\over\nu_h}\right)^{-p/2},~~~~~~~~~~~~~~~~\nu_h<\nu<\nu_{{\rm
KN},i},\\
\left({\nu_h\over\nu_l}\right)^{-(q-1)/2}\left({\nu_{{\rm KN},i}\over\nu_h}\right)^{-p/2}\left({\nu\over\nu_{{\rm KN},i}}\right)^{-(p-1)},~~~~\nu_{{\rm KN},i}<\nu,\\
\end{array}\right.\label{SSCspe}
\end{equation}
where $\nu_{l}=\min(\nu_{m,i}^{\dag},\nu_{c,i}^{\dag})$,
$\nu_{h}=\max(\nu_{m,i}^{\dag},\nu_{c,i}^{\dag})$, and $q=2$ for
$\nu_{c,i}^{\dag}<\nu_{m,i}^{\dag}$ and $q=p$  for
$\nu_{c,i}^{\dag}>\nu_{m,i}^{\dag}$.

Using equation (\ref{SSCspe}), we calculate the peak flux of GeV
$\gamma$-ray flares as shown by the shaded contours in Figure 5.
Comparing these GeV emission contours with the dashed contours that
correspond to the X-ray luminosity, a positive correlation between
these two emission components can be found. For the GRBs at a
typical distance of $10^{28}$cm, the high energy counterparts of the
relatively brighter X-ray flares could be detected by the LAT,
whereas those associated with weaker X-ray flares leak. The LAT
sensitivity here ($\rm few\times10^{-9}~erg~s^{-1}~cm^{-2}$) is
calculated for a typical flare onset time of several hundred seconds
and represented by the dash-dotted line in the figure. Furthermore,
according to the expression of $Y_i\approx
[(4\eta_i\epsilon_{e}/\epsilon_{B}+1)^{1/2}-1]/2$, we know that the
relative importance of the SSC and synchrotron emissions may be
sensitive to the parameters $\epsilon_{B}$ and $\epsilon_e$.
Therefore, we show the GeV $\gamma$-ray flux at the time of $T_{\rm
exp}$ varying in the ($\epsilon_{e}$, $\epsilon_B$) parameter space
in Figure 6. It can be seen that the high energy flux is mainly
sensitive to $\epsilon_e$ and equipartition values for $\epsilon_e$
are required. In addition, we would like to show some example
spectra numerically in Figure 7 using a more elaborate code that was
developed in Yu et al. (2007).

\section{Summary and discussion}
The LIS model is usually employed to explain the observed GRB
afterglow X-ray flares. However, a careful description for LIS
dynamics and some observational constraints on it still need to be
investigated. Based on this consideration, we studied the properties
of LIS-produced emission in the framework of a simplified paradigm,
i.e. internal forward-reverse shocks generated by a collision
between two homogeneous shells. With a lower limit for the observed
X-ray luminosity, we found a mildly high ratio of the Lorentz
factors of pre-collision shells, which leads to at least one
mildly-relativistic internal shock. Our results also show that the
brightest X-ray flares might imply a high variability of Lorentz
factors as indicated by $\gamma_4={\rm few}\times\gamma_1^2$. The
luminosity of the X-ray flares is mainly determined by the kinetic
energy luminosity of the delayed rapid shell rather than the leading
slow shell. After an investigation of the characteristic frequencies
in a wide parameter space, three types of theoretical X-ray light
curves are found, all of which are ended by a very steep decay with
$\alpha\sim p+1$. This indicates that the flare emission during the
decay phase is probably dominated by the curvature effect, which is
consistent with the observational inference found by Liang et al.
(2006).

We also investigated the peak flux of the GeV $\gamma$-ray
counterparts in the SSC scenario. By comparing the GeV flux with the
flux sensitivity of the \textit{Fermi} LAT, we found that the high
energy flares associated with relatively brighter X-ray flares could
be detected by the LAT for a distance of $10^{28}$cm to the source,
where an equipartition value of $\epsilon_{e}$ is required. This
possible detection will be very helpful to discriminate different
origins of high energy flares and different models for X-ray flares.
As mentioned above, two types of high energy flares are predicted by
the LIS model including the ones due to the SSC and EIC emission. In
the SSC case, a good temporal correlation between the X-ray and high
energy flares are expected, whereas a significant temporal extension
appears for high energy flares in the EIC case (Fan et al. 2007). In
addition, besides the LIS model, some authors suggested that X-ray
flares may be produced by a delayed external shock and the
corresponding high energy emission was also expected (Galli \& Piro
2007). But for the delayed afterglow model it is difficult to
explain the reoccurrence of X-ray flares in one GRB afterglow.

\section*{Acknowledgements}
This work is supported by the National Natural Science Foundation of
China (grants no. 10221001, 10640420144, and 10873009) and the
National Basic Research Program of China (973 program) No.
2007CB815404. YWY is also supported by the Scientific Innovation
Foundation of Huazhong Normal University, the Visiting PhD Candidate
Foundation of Nanjing University and the National Natural Science
Foundation of China (grant 10773004).

\begin{table}
\caption{Four possible relationships (a, b, c and d) between $\nu_{m}^*$, $\nu_c^*$, and $\nu_X$ and three types
(I$-$III) of the corresponding X-ray flare light curves (characterized by the temporal indices).}
\begin{center}
\begin{tabular}{c|c|c}\hline\hline
Regimes                              & Types  &      Temporal Indices ($\alpha$)        \\
 \hline
a: $\nu_m^*<\nu_X<\nu_c^*$           & I      &  $-1, {sp+3\over3}$                                                \\
b: $\nu_m^*<\nu_c^*<\nu_X$           & II     &  $-1, 0, {sp-2s+3\over3}, {sp+3\over3}$                            \\
c: $\nu_c^*<\nu_m^*<\nu_X$           & II     &  $-1, 0, {sp-2s+3\over3}, {sp+3\over3}$                            \\
d: $\nu_c^*<\nu_X<\nu_m^*$           & III    &  $-1, -{5\over3}, 0, {3-s\over3}, {sp-2s+3\over3}, {sp+3\over3}$   \\
 \hline
 \hline
\end{tabular}
\end{center}
\end{table}
%
\begin{figure}\resizebox
{0.5\textwidth}{!} {\includegraphics{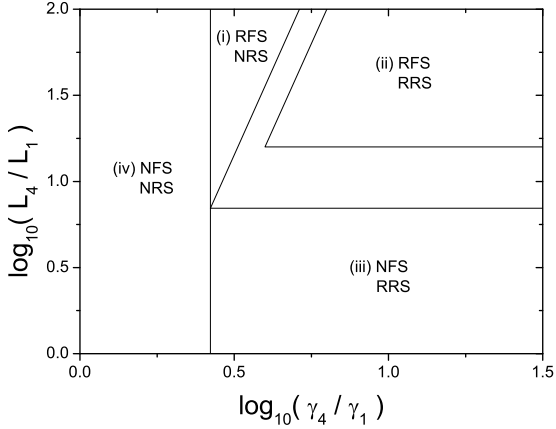}} \caption{Regions in
the ($\gamma_4/\gamma_1$, $L_4/L_1$) parameter space where four
limits of the LIS dynamics are given.}
\end{figure}
%
%
\begin{figure}
\resizebox{\hsize}{!}
{\includegraphics{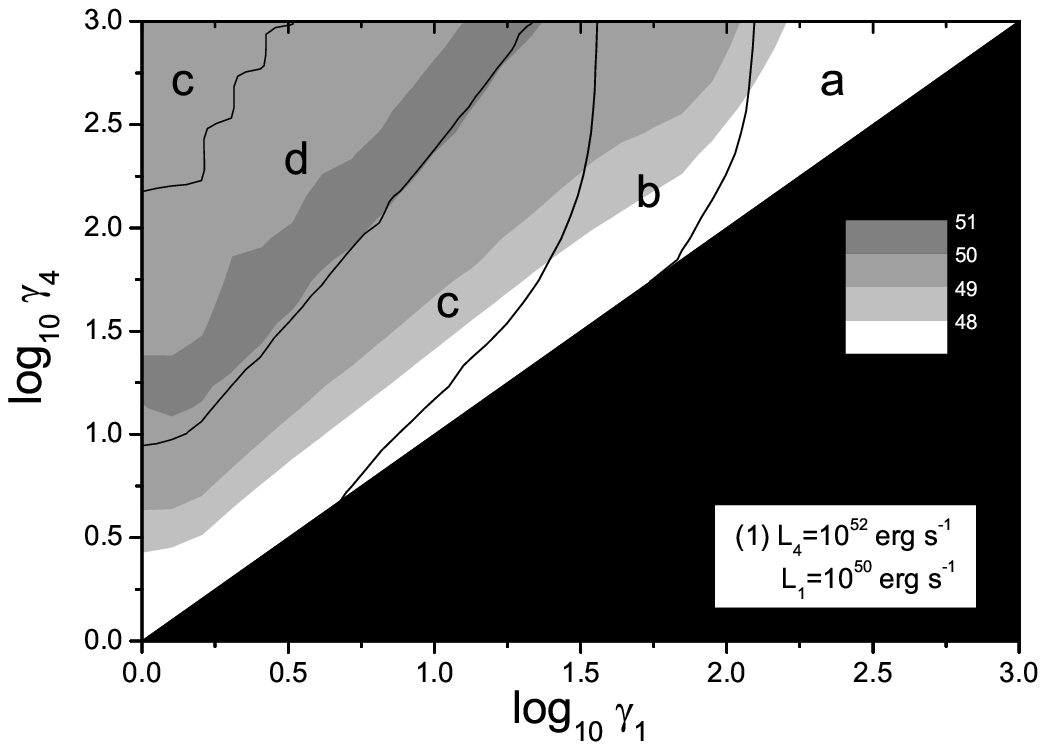}\includegraphics{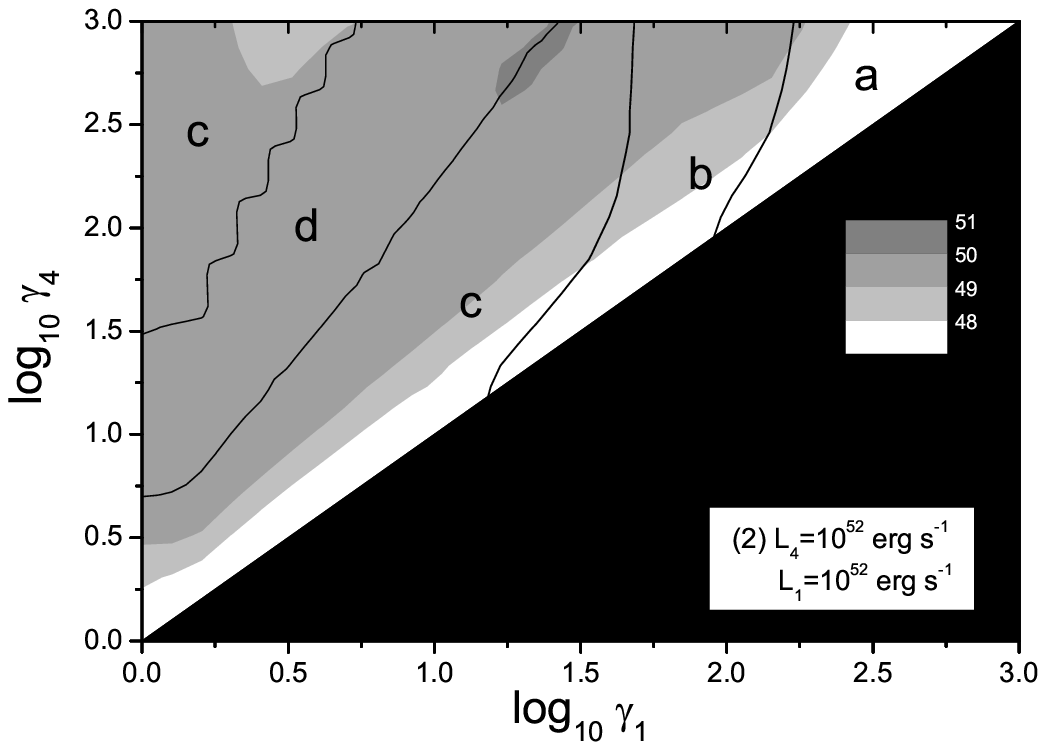}}
\caption{X-ray luminosity in the ($\gamma_1$, $\gamma_4$) parameter
spaces. The regions where the model-predicted X-ray luminosity
exceeds the observational lower limit ($\sim10^{48}\rm erg~s^{-1}$)
are presented by shaded contours, while the unshaded region is ruled
out by this luminosity constraint. Separating by solid lines, the
parameter spaces are divided into different regions denoted by ``a,
b, c, d", where different relationships between $\nu_c^*$,
$\nu_m^*$, and $\nu_X$ are given as listed in Table 1. The fixed
values of the kinetic-energy luminosities of the shells satisfy
$L_4/L_1\gg1$ for the left panel and $L_4/L_1=1$ for the right
panel, and their corresponding dynamic cases can be found in Figure
1. The black region is forbidden due to $\gamma_4<\gamma_1$. The
other model parameters $\epsilon_{B}$, $\epsilon_{e}$, $p$, and
$\Delta t_{\rm ej}$ are taken to be typical values of 0.03, 0.3,
2.5, and 100s, respectively.}
\end{figure}
%
%
\begin{figure}
\resizebox{\hsize}{!}
{\includegraphics{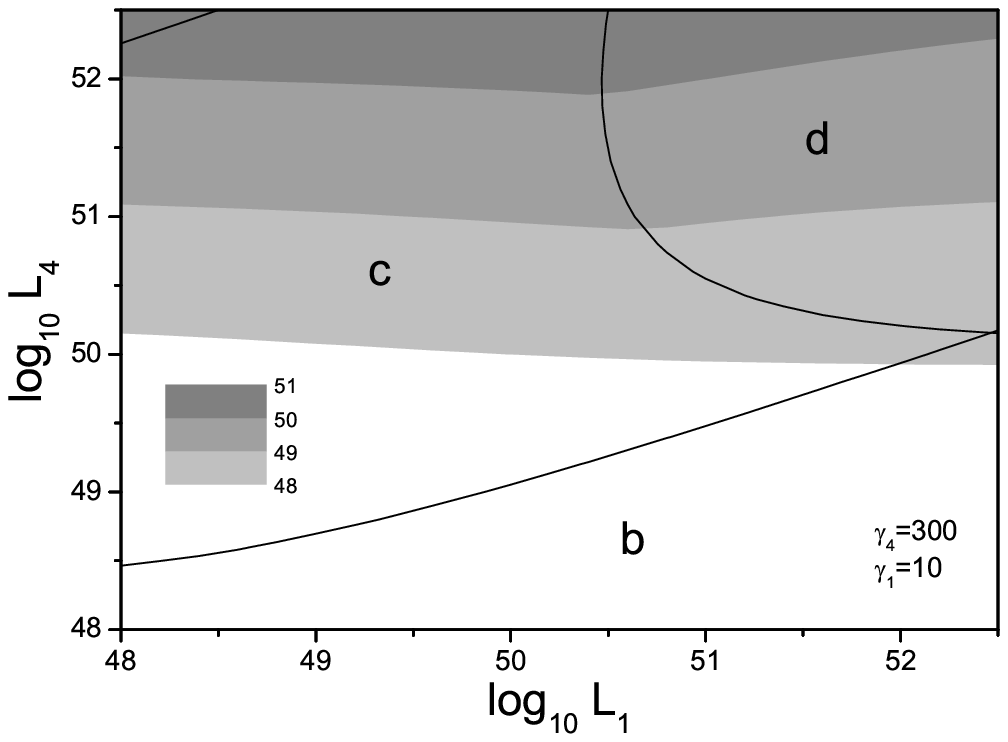}\includegraphics{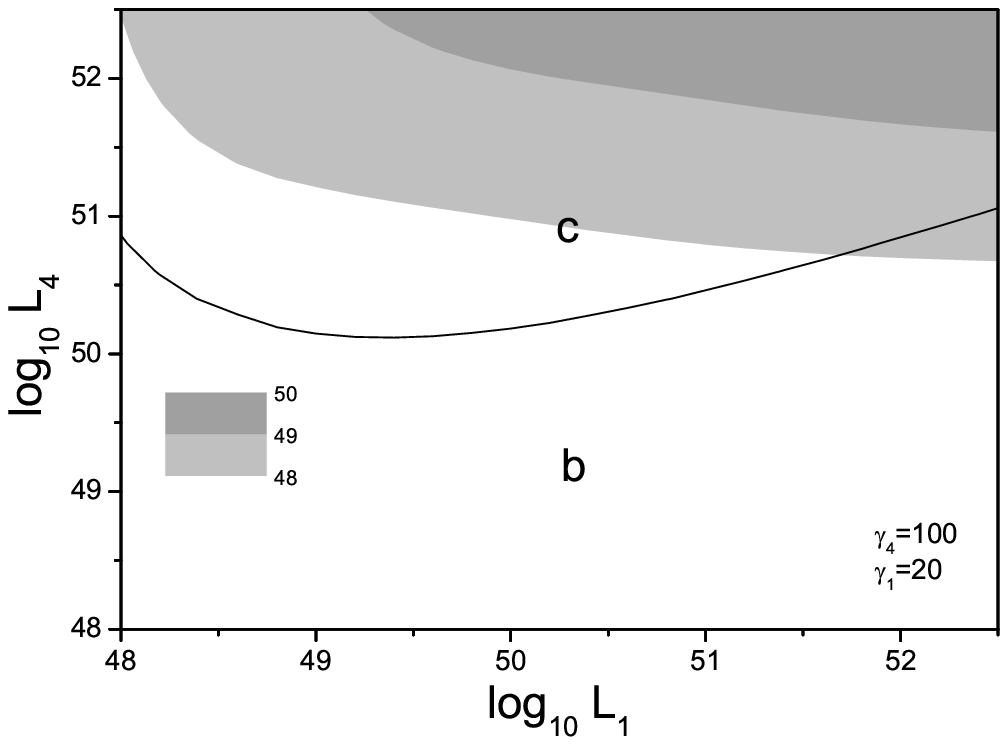}}
\caption{X-ray luminosity in the ($L_1$, $L_4$) parameter spaces.
The left and right panels correspond to relatively higher and lower
values of the ratio $\gamma_4/\gamma_1$, respectively. The meanings
of the regions and the other model parameters are the same as those
in Figure 2.}
\end{figure}
%
%
\begin{figure}
\resizebox{0.5\textwidth}{!} {\includegraphics{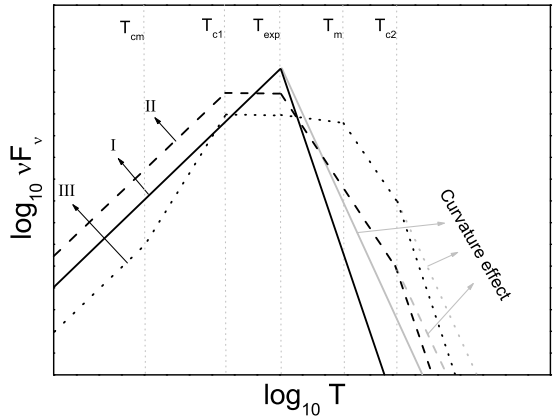}}
\caption{Schematic illustration of theoretical X-ray flare light
curves for $T_{\rm crs}=T_{\rm exp}$ (black lines; the temporal
indices of all segments are listed in Table 1). The curvature effect
is exhibited by the grey lines. The vertical dotted lines represent
the break times of the light curves, specifically, the time $T_{\rm
cm}$ at which $\nu_c=\nu_m$, the time $T_{\rm m}~(T_{\rm c})$ at
which the break frequency $\nu_m~(\nu_c)$ passes through the X-ray
band, the time $T_{\rm exp}$ from which the radius increases
linearly.}
\end{figure}
%
%
\begin{figure}\resizebox{0.5\textwidth}{!} {\includegraphics{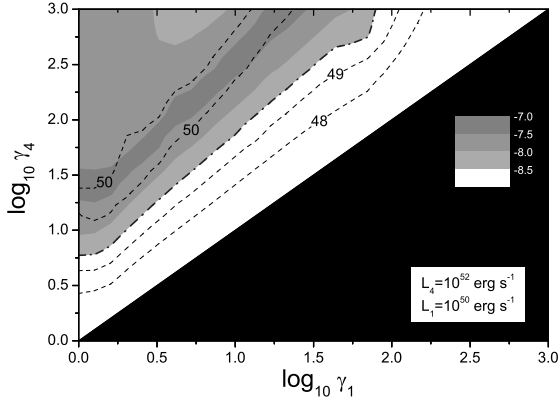}}
\caption{The peak flux at $T=T_{\rm exp}$ of GeV $\gamma$-ray flares
due to SSC emission for GRBs at the luminosity distance $10^{28}$cm
in the ($\gamma_1$, $\gamma_4$) parameter space. The regions where
the GeV flux exceeds the \textit{Fermi} LAT sensitivity (dash-dotted
line) are shown by shaded contours, while the SSC emission
calculated in the unshaded region could be not detected by the LAT.
The model parameters here are the same as those adopted in the left
panel of Figure 2. To compare with the associated X-ray component,
the X-ray luminosities are also shown by the dashed contours and
labeled by $\log_{10}[(\nu L_{\nu})_X/{\rm erg~s^{-1}}]$.}
\end{figure}
%
%
\begin{figure}\resizebox{0.5\textwidth}{!} {\includegraphics{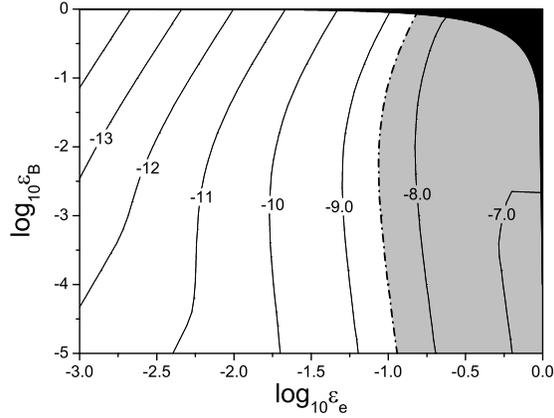}}
\caption{Variation of the peak flux at $T=T_{\rm exp}$ of GeV
$\gamma$-ray flares due to SSC emission for GRBs at the luminosity
distance $10^{28}$cm in the ($\epsilon_e$, $\epsilon_B$) parameter
space. The contours are labeled by the values of $\log_{10}[(\nu
F_{\nu})_{\rm GeV}/{\rm erg~s^{-1}cm^{-2}}]$. The region where the
GeV flux exceeds the \textit{Fermi} LAT sensitivity (dash-dotted
line) is shaded. The black region is forbidden due to
$\epsilon_B+\epsilon_e>1$. The other model parameters are taken to
be $L_1=10^{50}~\rm erg~s^{-1}$, $L_4=10^{52}~\rm erg~s^{-1}$,
$\gamma_1=10$, $\gamma_4=300$, $p=2.5$, $\Delta t_{\rm ej}=100$s and
$t_{\rm ej,A}=400$s.}
\end{figure}
%
%
\begin{figure}\resizebox{0.5\textwidth}{!} {\includegraphics{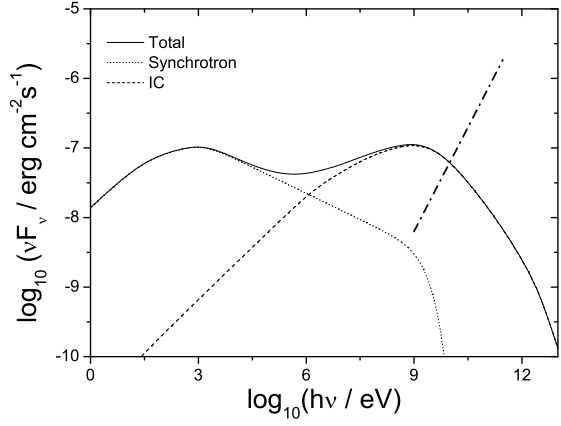}}
\caption{An example numerically-calculated spectrum by combining the
synchrotron and SSC spectra contributed by the two shocks for the
luminosity distance of $10^{28}$cm. The dash-dotted line denotes the
\textit{Fermi} LAT sensitivity. The model parameters are taken to be
$\epsilon_{B}=0.03$, $\epsilon_{e}=0.3$, $L_1=10^{50}~\rm
erg~s^{-1}$, $L_4=10^{52}~\rm erg~s^{-1}$, $\gamma_1=10$,
$\gamma_4=300$, $p=2.5$, $\Delta t_{\rm ej}=100$s and $t_{\rm
ej,A}=400$s.}
\end{figure}
%
\end{document}